# Higher-order Dirac sonic crystals

Huahui Qiu,[1] Meng Xiao,[1] Fan Zhang,[2] and Chunyin Qiu[1*]

[1]Key Laboratory of Artificial Micro- and Nano-Structures of Ministry of Education and School of Physics and Technology, Wuhan University, Wuhan 430072, China

[2]Department of Physics, University of Texas at Dallas, Richardson, Texas 75080, USA

**Abstract:** Discovering new topological phases of matter is a major theme in fundamental physics and materials science. Dirac semimetal provides an exceptional platform for exploring topological phase transitions under symmetry breaking. Recent theoretical studies have revealed that a three-dimensional Dirac semimetal can harbor fascinating hinge states, a higher-order topological manifestation not known before. However, its realization in experiment is yet to be achieved. In this Letter, we propose a minimum model to construct a *spinless* higher-order Dirac semimetal protected by $C_{6v}$ symmetry. By breaking different symmetries, this parent phase transitions into a variety of novel topological phases including higher-order topological insulator, higher-order Weyl semimetal, and higher-order nodal-ring semimetal. Furthermore, for the first time, we experimentally realize this unprecedented higher-order topological phase in a sonic crystal and present an unambiguous observation of the desired hinge states via momentun-space spectroscopy and real-space visualization. Our findings may offer new opportunities to manipulate classical waves such as sound and light.

[*]Corresponding author. cyqiu@whu.edu.cn

*Introduction.*—Topological phases of matter have been one of the most active research fields since the discovery of topological insulators (TI) [1-3]. One fundamental feature of such fascinating phases is the bulk implication of protected boundary modes, dubbed bulk-boundary correspondence. Recently, tremendous efforts have been devoted to predicting higher-order (HO) topological phases that admit unconventional bulk-boundary correspondence [4-19]. Unlike the conventional (first-order) topological phases that feature gapless modes at boundaries of one dimension lower, i.e., co-dimension $d_c = 1$, the hallmark of HO topological phases is the existence of gapless modes at boundaries of co-dimension $d_c > 1$, e.g., corner modes of a two-dimensional (2D) system and hinge modes of a three-dimensional (3D) system. In understanding such HO topological phases and their topological responses, spatial symmetries often play a critical role, in addition to time-reversal and/or particle-hole symmetries.

Nontrivial HO topology has been predicted not only in gapped insulators [4-9] but also in gapless semimetals [10-19]. These new HO members greatly enrich the already diverse spectrum of topological phases of matter. Interestingly, for a HO Dirac semimetal [10,13-17], its gapped 2D slices may be classified into two HO topologically distinct insulators, for which the transition occurs exactly at the Dirac points (DPs). Similar topological transitions also emerge in HO Weyl semimetals [11,12,18,19]. Besides the four-fold-degenerate DPs or two-fold-degenerate Weyl points for the linearly crossed bulk bands, such HO characteristics lead to the presence of fascinating one-dimensional (1D) hinge modes connecting the projected Dirac or Weyl nodes, a robust manifestation of the HO bulk-hinge correspondence. This hallmark is markedly different from the surface Fermi arcs previously observed in conventional topological semimetals [3,20-24].

To date there have been extensive studies of various HO topological phases in condensed matter and materials physics [4-45], ranging from solid-state materials to ultra-cold atoms [4-9], from photonics [26,28,32-35,45] to acoustics [29-31,39-44], and from mechanics [25,36] to electro-circuits [27,37,38]. In particular, the classical artificial crystals have been demonstrated to be exceptional platforms for exploring those otherwise elusive topological phases, benefiting from their macroscopically more controllable structures and less demanding measurements [46-49]. A variety of HO TIs have been ingeniously realized in such classical systems soon after their theoretical predictions [25-45]. However, little experimental progress has been made toward the HO topological nodal phases [50,51], especially for the Dirac semimetal phase (despite that a family of *spin-orbit-coupled* solid-state materials has been suggested [16]). In this Letter, we theoretically propose a simple scheme to construct a *spinless* HO Dirac semimetal, as a parent phase for a novel family of HO topological phases including TIs, Weyl semimetals, and nodal-ring semimetals, experimentally realize it in a 3D Dirac sonic crystal (DSC), and unambiguously present the first observation of its fascinating hinge modes.

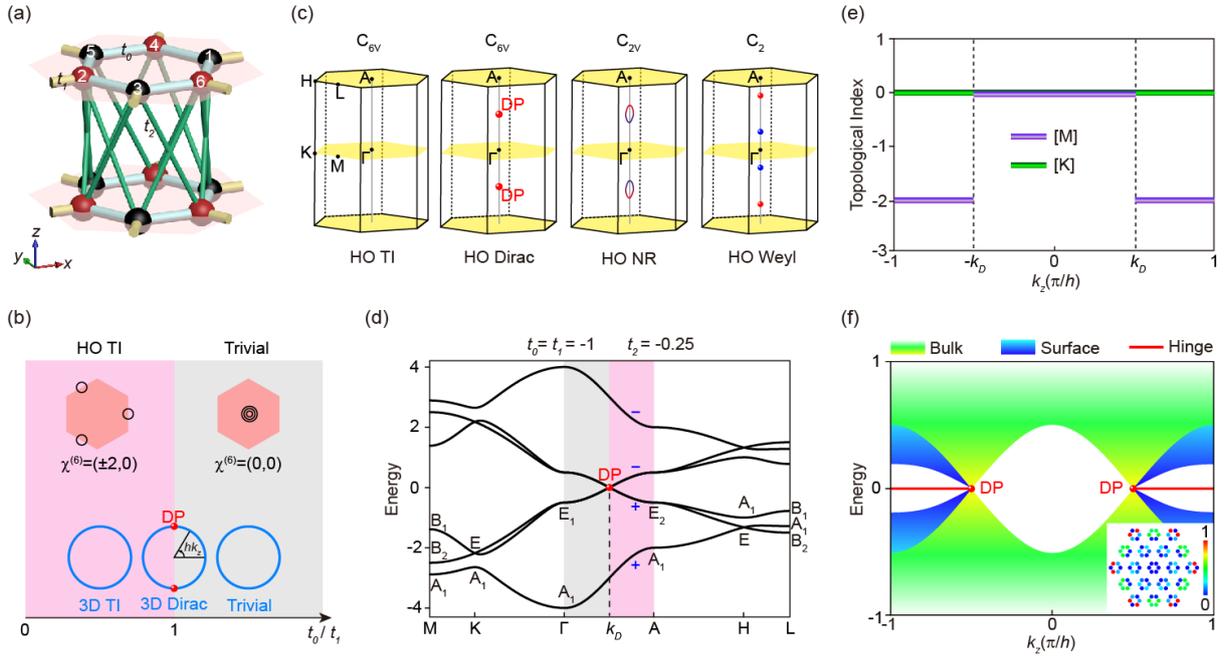

FIG. 1. Tight-binding model for constructing 3D HO topological phases. (a) 3D tight-binding model. The six orbitals in each unit cell form two different sublattices (colored spheres). $t_0$ is the intra-cell coupling, and $t_1$ and $t_2$ are the in-plane and out-of-plane inter-cell couplings. (b) Reduced topological phase diagram for the 2D monolayer system. The trivial and HO TI phases have different topological indices $\chi^{(6)} = ([M], [K])$ and Wannier centers (black circles) for the lower three bands. The blue circles sketch three possible evolution scenarios of the $k_z$-slices when $hk_z$ varies from $-\pi$ to $\pi$. (c) Multiple 3D HO topological phases realized by our model, including TI and Dirac, Weyl, and nodal-ring (NR) semimetals. The colored points (rings) label the HO Dirac or Weyl points (NRs) that come in pairs. (d) Bulk band structure for a set of couplings that features HO DPs. The bands are labeled by the $C_{6v}$ irreducible representations at high-symmetry momenta and the $C_2$ parities along the ΓA line. (e) Topological indices $\chi^{(6)} =$ ([M], [K]) plotted as a function of $k_z$. (f) Hinge-projected spectrum calculated for an infinitely long hexagonal prism. It features zero-energy hinge modes (red lines) connecting the projected DPs. Inset: Typical in-plane eigenfields of the hinge modes.

*Tight-binding models.*—We first present in Fig. 1(a) a minimal tight-binding model for our designed 3D system that can exhibit multiple HO topological phases. Each unit cell includes six spinless orbitals with intra-cell couplings $t_0$, inter-cell couplings $t_{1,2}$, and vanishing onsite energies. The in-plane and out-of-plane lattice constants are $a$ and $h$, respectively. Each monolayer forms a triangular latttice, i.e., a $\sqrt{3} \times \sqrt{3}$ superstructure of an original hexagonal latttice, and can be modeled by a Hamiltonian $H_{2D} = t_0 H_0 + t_1 H_1$. $H_0$ is a constant matrix with its nontrivial entries representing the nearest-neighbour couplings within the same unit cell. $H_1(\boldsymbol{k}_\parallel) = \oplus_{i=1}^3 [\cos(\boldsymbol{k}_\parallel \cdot \boldsymbol{a}_i)\sigma_x - \sin(\boldsymbol{k}_\parallel \cdot \boldsymbol{a}_i)\sigma_y]$ corresponds to couplings between the neighbouring unit cells, where $\boldsymbol{k}_\parallel$ is the in-plane Bloch wavevector, $\sigma_{x,y}$ are the Pauli matrices coupling the two sublattices, and $\boldsymbol{a}_i$ are the 2D primitive lattice vectors $\boldsymbol{a}_1 = (a, 0)$ and $\boldsymbol{a}_{2,3} = (-\frac{a}{2}, \mp\frac{\sqrt{3}a}{2})$. In additon to the $C_{6v}$ and time-reversal symmetries, $H_{2D}$ enjoys a chiral symmetry since the six orbitals lie in two decoupled sublattices. Recently, it has been demonstrated [28,52] that the nontrivial HO topology of such a 2D insulator with $C_6$ symmetry can be characterized by two topological invariants $\chi^{(6)} =$ ([M], [K]), where the index [M] (or [K]) is a measure of the difference in $C_2$ (or $C_3$) representation for the valence-band states at the high-symmetry momenta Γ and M (or K). Specifically, the $C_2$-invariant [M] ∈ Z is defined as [M] = #$M_1$ − #$\Gamma_1^{(2)}$, where #$M_1$ and #$\Gamma_1^{(2)}$ count the numbers of negative energy states with $C_2$ eigenvalue +1 at M and Γ, respectively. Similarly, the $C_3$-invariant [K] ∈ Z is defined as [K] = #$K_1$ − #$\Gamma_1^{(3)}$, where #$K_1$ and #$\Gamma_1^{(3)}$ are the numbers of negative energy states with $C_3$ eigenvalue +1 at K and Γ, respectively. Figure 1(b) shows a reduced phase diagram for our monolayer model. Whereas the $C_3$ symmetry operator commutes with the chiral symmetry operator yielding a trivial $C_3$-invariant $[K] = 0$ [28,52], the $C_2$ symmetry relates the two different sublattices and can give rise to nontrivial HO band topology. For $0 < t_0/t_1 < 1$, the $C_2$-invariant $[M] = \pm 2$, where $\pm$ denote the signs of $t_1$. In sharp contrast to the trivial phase (i.e., $[M] = 0$ for $t_0/t_1 > 1$) with Wannier centers at the unit cell center, the nontrivial phase with Wannier centers at the sides of the unit cell exhibits zero-energy corner modes in a finite-sized sample (*see Supplementary Information*). In particular, at the phase boundary $t_0/t_1 = 1$, the bulk gap closes and the 2D system features a fourfold degenerate DP at the $\Gamma$ point, as an accidental degeneracy between two inequivalent 2D irreducible $C_{6v}$ representations; the two double degeneracies at the K and K′ points of the original hexagonal lattice are folded into the $\Gamma$ point of its $\sqrt{3} \times \sqrt{3}$ triangular superlattice.

Our 3D system can also be modeled by the Hamiltonian $H_{2D}$ if $t_0$ is replaced by $t_{eff} = t_0 + 2t_2 \cos(hk_z)$, given that the two extra $z$-directed couplings (each valued $t_2$) play similar roles to the in-plane coupling $t_0$. As $hk_z$ varies from $-\pi$ to $\pi$, the relative intra-cell to inter-cell coupling $t_{eff}/t_1$ experiences a circle of radius $|2t_2/t_1|$ around the origin $(t_0/t_1, 0)$, as illustrated in Fig. 1(b). When the origin is in the nontrivial phase side and the radius is sufficiently small, e.g., the HO TI monolayers are barely coupled, any $k_z$ slice realizes the 2D HO TI phase, and the stacked system is a 3D HO TI (with perfectly flat zero-energy hinge states due to the chiral symmetry). On the contrary, the stacked system is topologically trivial when the origin is in the trivial phase side and the radius is sufficiently small. Intriguingly, when the circle traverses the phase boundary, one time-reversal pair of DPs emerge in the rotational axis, as illustrated in Fig. 1(c). To examplify this Dirac semimetal phase, we consider the case with $t_0 = t_1 = -1$ and $t_2 = -0.25$. Figure 1(d) plots the bulk band structure that features one DP at $k_D = \frac{\pi}{2h}$ in the high-symmetry line $\Gamma$A. (The other DP is at $k_z = -k_D$.) As shown in Fig. 1(e), the $C_2$-invariant is nontrivial ($[M] = -2$) for $|k_z| > |k_D|$, while the $C_3$-invaraint $[K]$ vanishes everywhere. The nontrivial topological invariants $\chi^{(6)} = (-2, 0)$ implies the existance of hinge states connecting the hinge-projected DPs, since all the $k_z$-slices within $|k_z| > |k_D|$ host corner modes. This is further confirmed by the hinge spectrum of an infinitely long hexagonal prism in Fig. 1(f), where there exist six symmetric 1D hinge states at exactly zero energy (because of the chiral symmetry). Clearly, the derived Dirac semimetal, sharply different from the conventional Dirac semimetals, has a HO nature.

Remarkably, the HO DPs are protected by the $C_{6v}$ symmetry. In Fig. 1(d), the DP is an accidental crossing point in the ΓA line between two pairs of bands that have opposite $C_2$ parities, and each pair itself forms a 2D irreducible representation of the $C_{3v}$ point group. When the $C_3$ symmetry is broken, the DP evloves into a zero-energy nodal ring (NR) in the intact mirror plane. When both the $C_3$ and mirror symmetries are broken, the DP splits into two Weyl points in the ΓA line. We stress that all these 3D HO topological phases summarized in Fig. 1(c) are unprecedented. Evidently, the symmetry protections of the HO bulk nodes are sharply different from their conventional (first-order) counterparts. At boundary, while gapless surface states only exist for the HO Weyl and NR semimetals, characteristic hinge states do exist for all the four HO topological phases due to the presence of the nontrivial $C_2$-invariant. Hereafter, we will focus on the HO Dirac semimetal, since it is a parent phase for all the other HO topological phases listed in Fig. 1(c) (*see Supplementary Information*).

*Acoustic realization and characterization of the HO Dirac semimetals.*—We further emulate our 3D tight-binding model in an acoustic system, in which cavity resonantors and narrow tubes are used respectively to mimic orbitals and couplings [29,30]. In addition to identifying the bulk DPs, we confirm the HO nature by obtaining the $k_z$-resolved [M] index and the flat hinge states at the resonant frequency (*see Fig. S3 in Supplementary Information*). Importantly, the HO DPs can be realized in any acoustic structure that features the same space group and topological invariants, as shown numerically and experimentlaly in a more general DSC below. Comparing with the cavity-tube system, this design relaxes the chiral symmetry, which is inessential for protecting the HO topology, and possesses unique merits, such as broad bulk bands and dispersive hinge states (due to the absence of resonance) and convenient experimental detections (due to the improved structural porosity and connectivity).

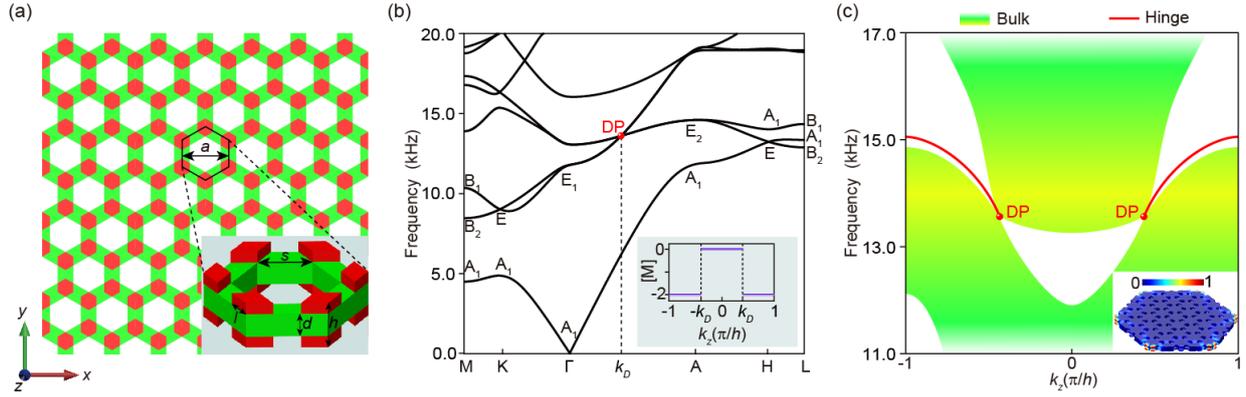

FIG. 2. Acoustic realization of the (type-II) HO DPs. (a) In-plane lattice and strcuture of our 3D DSC. (b) Bulk band structure along high-symmetry directions. At the high-symmetry points, the lowest three bands are characterized by the same irreducible representations to the lowest three bands in Fig. 1d, with a nontrivial [M] index for $|k_z| > k_D$ (inset). (c) Hinge-projected spectrum. Inset: Pressure amplitude distribution (color) of the hinge states of 14.86 kHz simulated in a small-size system. Notice that the colored region corresponds to the air domain of our HO DSC, which complements spatially with the colored rigid structure displayed in (a).

Figure 2(a) displays the structure of our layered DSC. The in-plane and out-of-plane lattice constants are $a = 27.7$ mm and $h = 9.6$ mm, respectively. Each layer is formed by a hard plate (of thickness $d = 4.8$ mm) perforated with a triangle lattice of hexgonal holes (of sidelength $s = 10.7$ mm). The layers are connected by hexgonal pillars of sidelength $l = 5.3$ mm. The whole plate-pillar structure [colored in Fig. 2(a)] is assumed to be acoustically rigid, and we consider the airborne sound in the remaining space. Our full-wave simulations are performed by using COMSOL Multiphysics (*see Supplementary Information*). Figure 2(b) plots the bulk band structure along high-symmetry lines. Clearly, there exists a fourfold degenerate DP at frequency ~13.7 kHz and $k_D \sim 0.4\pi/h$ along the ΓA line. Note that the Dirac cone, dubbed type-II [3,53,54], is strongly tilted since the chiral symmetry is not present. After inspecting the eigenfields of the lowest three bands at the high-symmetry points, we find that the acoustic system features the same irreducible representations to those of the tight-binding model in Fig. 1(d). As such, the $C_2$-invariant [M] is nontrivial for the momentum range $|k_z| > k_D$ [see the inset of Fig. 2(b)], which is suggestive of a hinge state connecting the time-reversal pair of projected DPs through $|k_z| > k_D$. This is indeed confirmed by the hinge spectrum simulated for a finite-sized sample, as shown in Fig. 2(c). Moreover, the inset of Fig. 2(c) shows the strong hinge localization of the hinge states. Note that the absence of chiral symmetry enables broadband hinge states that can be easily frequency-resolved in our airborne sound experiments below.

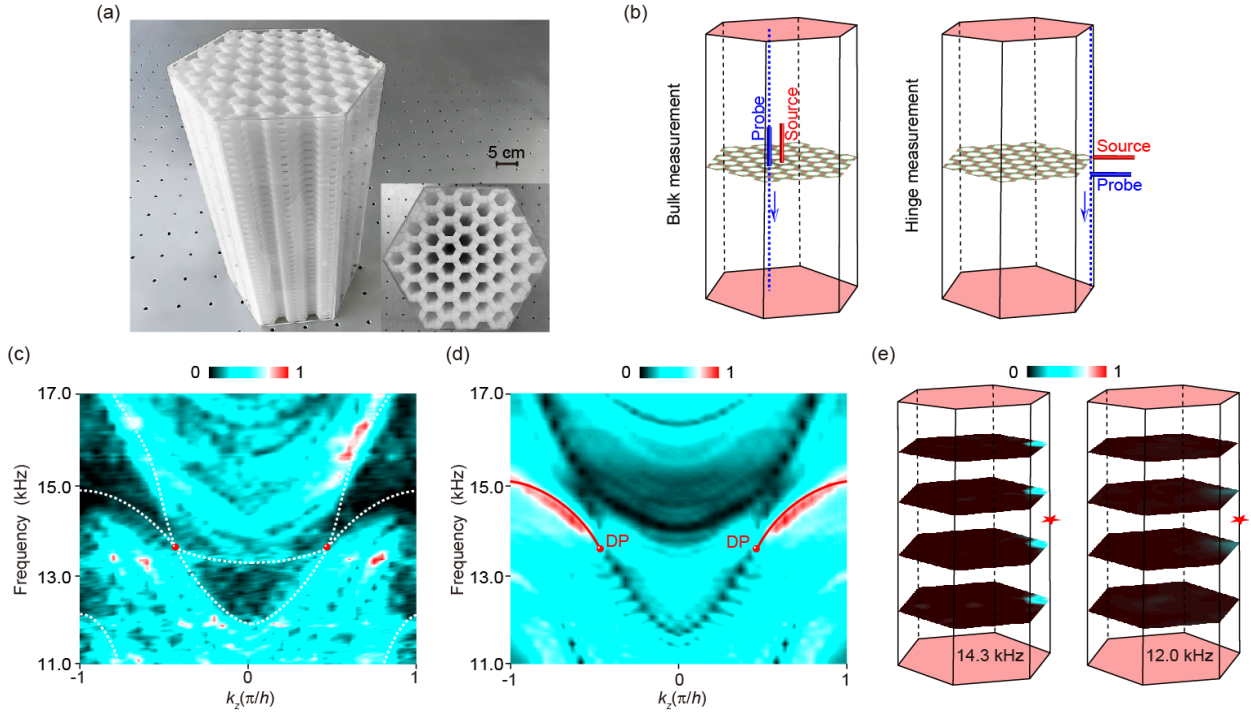

FIG. 3. Experimental observation of the HO DPs and 1D hinge states. (a) Side and top views of our sample. (b) Schematics of our bulk and hinge measurements. The blue dotted line indicates the scanning path in each case. (c) Bulk spectrum (color scale) excited by a sound source inserted into the middle of the sample interior. The white dashed lines sketch the boundaries of the projected bulk bands extracted from Fig. 2(c). (d) Hinge spectrum (color scale) excited by a sound source placed in the middle of the sample hinge, which reproduces well the simulated hinge states (red line) in Fig. 2(c). (e) Pressure amplitude distributions scanned in four equidistant planes of the sample, exemplified at 14.3 kHz and 12.0 kHz. The red star labels the position of the sound source.

Finally, we perform airborne sound experiments to conclusively identify the DPs and hinge states, the two defining signatures of the HO DSC. Figure 3(a) shows our experimental sample, fabricated precisely via 3D printing with photosensitive resin. It consists of 49 unit cells in the $x$-$y$ plane and 40 layers along the $z$ direction. Acrylic plates are closely pasted on the six side surfaces to mimic the hard boundary condition implemented in the simulation. We first measure the bulk bands of our DSC. As illustrated in the left panel of Fig. 3(b), we insert a point-like sound source into the middle of the sample from its top facet and scan the pressure distribution along a vertical hole adjacent to the sound source. Figure 3(c) presents the 1D Fourier transform of the measured data, with red and black indicating strong and weak amplitudes, respectively. As a whole, the observed bulk spectrum captures the full-wave simulated boundaries of the projected bulk bands in Fig. 2(c) and exhibits the expected band touching around each predicted DP.

Next, we measure the 1D hinge spectrum. We position the sound source in the middle of one hinge and scan the pressure distribution along the hinge (*see Supplementary Information*), as illustrated in the right panel of Fig. 3(b). Exhibiting as the brightest signals in Fig. 3(d), the dispersive hinge states emanate from the two projected DPs and match well with the full-wave simulation in Fig. 2(c). Essentially, the presence of 1D topological hinge states instead of 2D surface states is the HO manifestation of our DSC, which is markedly different from the conventional (first-order) Dirac semimetals with *surface* arc states [3,21,22]. In addition to the spectroscopy measurement, we also scan the pressure field pattern over the entire sample with the same hinge excitation. Figure 3(e) shows the pressure amplitude profiles in four equidistant planes of the sample at two typical frequencies. As expected, at 14.3 kHz a strong localization of the pressure field is observed around the sample hinge where the sound source is positioned. This forms a sharp contrast to the weak field distribution displayed at 12.0 kHz, a frequency away from the hinge states. Such extremely localized HO hinge modes are anticipated to have potential applications, such as acoustic sensing and energy trapping.

*Conclusions*. — we have proposed a series of unprecedented 3D HO topological phases with a simple scheme, in which the parent phase possesses $C_{6v}$ symmetry-protected DPs and hinge states. In particular, we have successfully fabricated a HO DSC and unambiguously identified its defining signatures, i.e., the DPs and hinge states, by performing full-wave simulations and airborne sound experiments. The design of our DSC enables us to observe not only the hinge states in frequency-resolved spectroscopy but also their spatial localization in pressure-field distributions. Not only is this HO DSC sharply different from the conventional solid-state Dirac semimetals that are characterized by surface Fermi arcs [3,21,22], but it is also markedly different from the recently achieved non-symmorphic DSCs that feature symmetry-enforced DPs and quad-helicoid surface arcs [55,56]. Last but not at least, our scheme for achieving spinless 3D HO topological phases can be readily generalized to other classical systems with similar space groups.

**Acknowledgements**

C.Q. is supported by the National Natural Science Foundation of China (Grant No. 11890701), the Young Top-Notch Talent for Ten Thousand Talent Program (2019-2022), and the Fundamental Research Funds for the Central Universities (Grant No. 2042020kf0209). M.X. is supported by the National Natural Science Foundation of China (Grant No. 11904264). F.Z. is supported by the UT Dallas Research Enhancement Fund.

**Author contributions**

C.Q. conceived the idea and supervised the project. H.Q. did the simulations and experiments. C.Q., H.Q., M.X., and F.Z. analyzed the data and wrote the manuscript. All authors contributed to scientific discussions of the manuscript.

*Supplementary Information for*

# Higher-order Dirac sonic crystals

Huahui Qiu,[1] Meng Xiao,[1] Fan Zhang,[2] and Chunyin Qiu,[1]*

[1]Key Laboratory of Artificial Micro- and Nano-structures of Ministry of Education and School of Physics and Technology, Wuhan University, Wuhan 430072, China

[2]Department of Physics, University of Texas at Dallas, Richardson, Texas 75080, USA

# 1. Higher-order (HO) topology in a 2D system with $C_6$ symmetry

Here we briefly review the HO topology of 2D systems with $C_6$ symmetry [1,2], which can be diagonsed by scrutinizing the irreducible group representations of the bands at high-symmetry points in the 2D hexgonal Brillouin zone, i.e., $\Gamma$, M, and K. Specifically, the HO topology of the system can be classified by the topological index $\chi^{(6)} = ([M], [K])$, in which the component [M] (or [K]) is a measure of the difference in the $C_2$ (or $C_3$) representations at $\Gamma$ and M (or K). The $C_2$-invariant $[M] \in Z$ is defined as $[M] = \#M_1 - \#\Gamma_1^{(2)}$, where $\#M_1$ and $\#\Gamma_1^{(2)}$ count the numbers of negative energy states with $C_2$ eigenvalue +1 at M and $\Gamma$, respectively. Similarly, the $C_3$-invariant $[K] \in Z$ is defined as $[K] = \#K_1 - \#\Gamma_1^{(3)}$, where $\#K_1$ and $\#\Gamma_1^{(3)}$ are the numbers of negative energy states with $C_3$ eigenvalue +1 at K and $\Gamma$, respectively. The system is topologically nontrivial for a nonzero $\chi^{(6)}$, which is invariant without closing the bulk band gap or breaking the $C_6$ symmetry. Physically, a nontrivial topological index leads to a fractional charge at the corners of a finite-sized sample: $Q_c = \frac{e}{4}[M] + \frac{e}{6}[K] \mod e$, which is a HO manifestation as zero-energy corner modes in the presence of chiral symmetry.

Previous studies [1,2] show that a topological phase transition occurs at $t_0/t_1 = 1$, where the bulk gap closes at $\Gamma$ point. Specifically, the gapped system is topologically trivial for $t_0/t_1 > 1$, with the bulk index $\chi^{(6)} = (0,0)$ and the coner charge $Q_c = 0$. The system becomes topologically nontrivial for $0 < t_0/t_1 < 1$, with the bulk index $\chi^{(6)} = (\pm 2, 0)$ and the corner charge $Q_c = e/2$. Here $\pm$ in the $C_2$-invariant [M] denote the signs of the inter-cell coupling $t_1$. Note that in the presence of chiral symmetry the $C_3$-invariant [K] is always zero since the $C_3$ operator commutes with the chiral operator [1]. This has also been checked by our calculations.

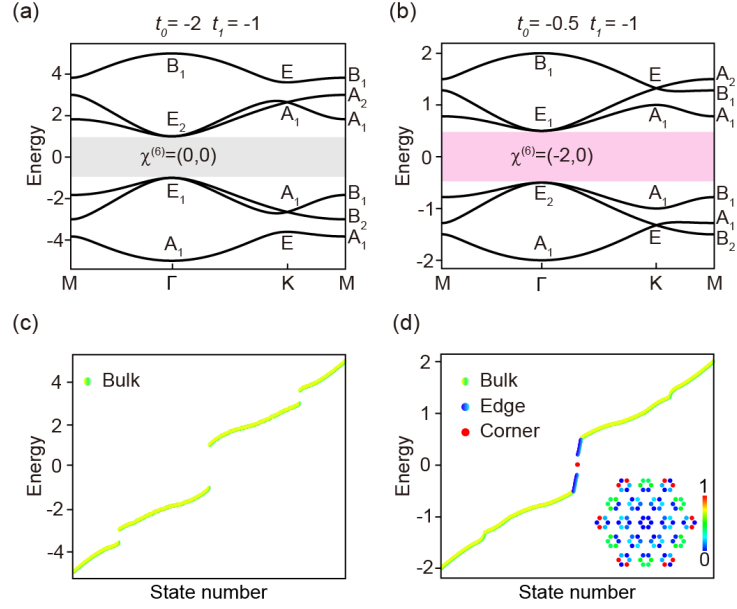

FIG. S1. 2D HO topological insulators with a nontrivial $\chi^{(6)}$ index. (a) and (b): Bulk band structures calculated for two different sets of couplings. Irreducible group representations are labeled at the high-symmetry momenta, which give trival and nontrivial $\chi^{(6)}$ indices for these two systems. (c) and (d): The corresponding corner spectra calculated for finite-sized samples. In contrast to the trival case, the corner spectrum of the nontrivial case exhibits robust corner modes pinned at zero energy (see inset), in addition to the trivial (gapped) edge modes.

Figure S1 presents two examples to show the HO band topology inherent in the 2D tight-binding model, one for the trvial case ($t_0 = -2$ and $t_1 = -1$) and the other for the nontrivial case ($t_0 = -0.5$ and $t_1 = -1$). After inspecting the eigenstates at the high-symmetry momenta of the three negative energy bands, we can determine the irreducible representations of the $C_6$ point group and identified the trivial and nontrivial $\chi^{(6)}$ indices for both systems. In contrast to the trivial case, the corner spectrum of the nontrivial case shows clearly robust corner modes pinned at zero energy.

## 2. Novel 3D HO topological phases predicted by our tight-binding model

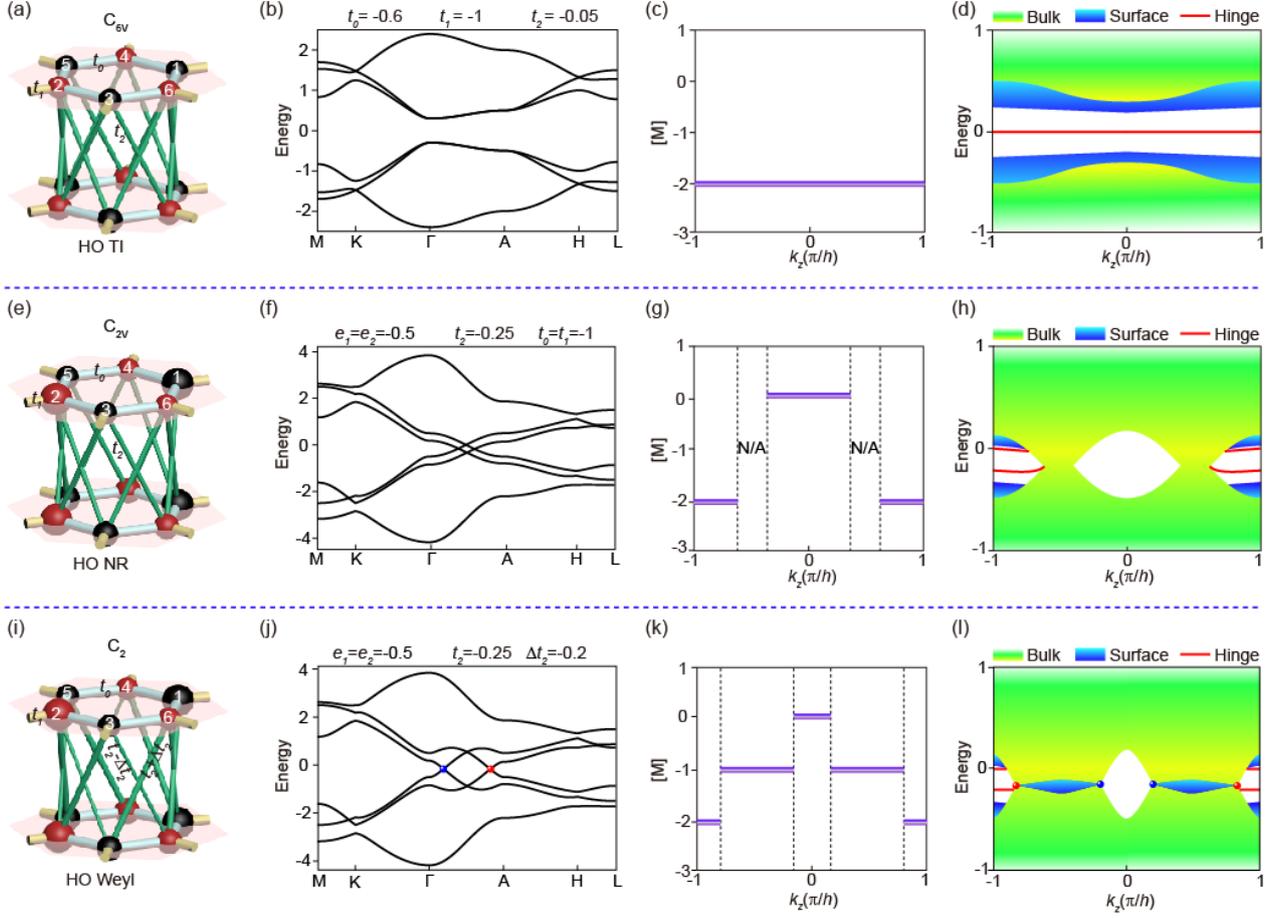

FIG. S2. Novel 3D HO topological phases predicted by our tight-binding model. We focus on the three 3D HO topological phases not discussed in detail in our main text, i.e., HO TI [(a)-(d)], HO nodal-ring (NR) semimetal [(e)-(h)], and HO Weyl semimetal [(i)-(l)]. (a), (e), and (i) sketch their tight-binding models, (b), (f), and (j) plot their bulk band structures, (c), (g), and (k) display the $k_z$-dependent $C_2$-invariants, and (d), (h), and (l) present their hinge spectra. For brievity, the hinge (or surface) states mixing with the projected bulk bands are not presented. Note that the irreducible Brillouin zone changes for the systems of different symmetries. For consistence, the high-symmetry momentum lines used for plotting the band structures remain the same as those used in Fig. 1d (see the main text).

Here we present some numerical details for the three 3D HO topological phases not discussed in detail in our main text [see Fig. 1(c)], including the HO TI [Figs. S2(a)-S2(d)], HO nodal-ring semimetal [Figs. S2(e)-S2(h)], and HO Weyl semimetal [Figs. S2(i)-S2(l)]. As shown in Fig. S2(a), the HO TI preserves the $C_{6v}$ symmetry but has different couplings compared with the HO Dirac semimetal in the main text. (Again, the two separated doubly degenerate bands along the ΓA line are enforced by the $C_{3v}$ symmetry). In this case, the hinge spectrum exhibits zero-energy hinge states across the entire hinge Brillouin zone [see Fig. S2(d)], since the same nontrivial $C_2$-invariant [M] $= -2$ persists for all $k_z$ [see Fig. S2(c)].

To realize the HO nodal-ring semimetal, we retain the couplings used in the HO Dirac semimetal but add nonzero onsite energies for the sites 1 and 2 ($e_1 = e_2 = -0.5$) to break the $C_3$ symmetry. In this case, the two double degeneracies in the ΓA line are lifted [see Fig. S2(f)], and each HO Dirac point evolves into a HO nodal ring in the ΓAK mirror plane [see Fig. 2(c) in the main text]. Figure S2(g) shows the $k_z$-dependent [M] index for this $C_2$-preserved HO nodal-ring system. (Note that there are three inequivalent M-points for each $k_z$ plane, which preserve $C_2$ rotation symmetry. We have calculated the $C_2$-invariants (at $k_z$ with gapped bulk bands) for all the three points and find them all trivial or all nontrivial simultaneously.) Comparing with the HO Dirac system [see Fig. 1(e) in the main text], the $k_z$ distribution of [M] persists outside the $k_z$ region projected by the gapless nodal-ring, in which the $C_2$-invariant is not well defined. The nontrivial $C_2$-invariant gives rise to the HO hinge states in Fig. S2(h), which are no longer pinned at zero energy because of the broken chiral symmetry. The hinge bands split in energy since the six hinges are inequivalent. In addition, the nontrivial surface states responsible for the first-order band topology of nodal rings, which mix with the projected bulk states, are not presented here for brevity. Similarly, the hinge states hidden in the projected bulk (or surface) bands are not displayed, either.

We further break the mirror symmetry in the HO nodal-ring system by assigning different strengths for the clockwise and anticlockwise interlayer couplings ($t_2 \pm 0.2$ used here). Each HO nodal ring evolves into one pair of Weyl points of opposite chirality along the $k_z$ axis [see Fig. 1(c) in the main text]. The hinge states are robust to the perturbation [see Fig. S2(l)], as the hallmark of the HO Weyl semimetal. Interestingly, hybrid band topology occurs within the newly gapped $k_z$ slices between the pair of Weyl points originating from the same Dirac point. The first-order topology is manifested as the nontrivial surface arcs within the corresponding $k_z$ regions [Fig. S2(l)]. The HO band topology, related to the nontrivial bulk invariant $[M] = -1$ in Fig. S2(k), does not yield well-defined hinge states because of their hybridization with bulk states or surface arcs.

### 3. Acoustic emulation of the HO Dirac semimetal predicted by our 3D tight-binding model

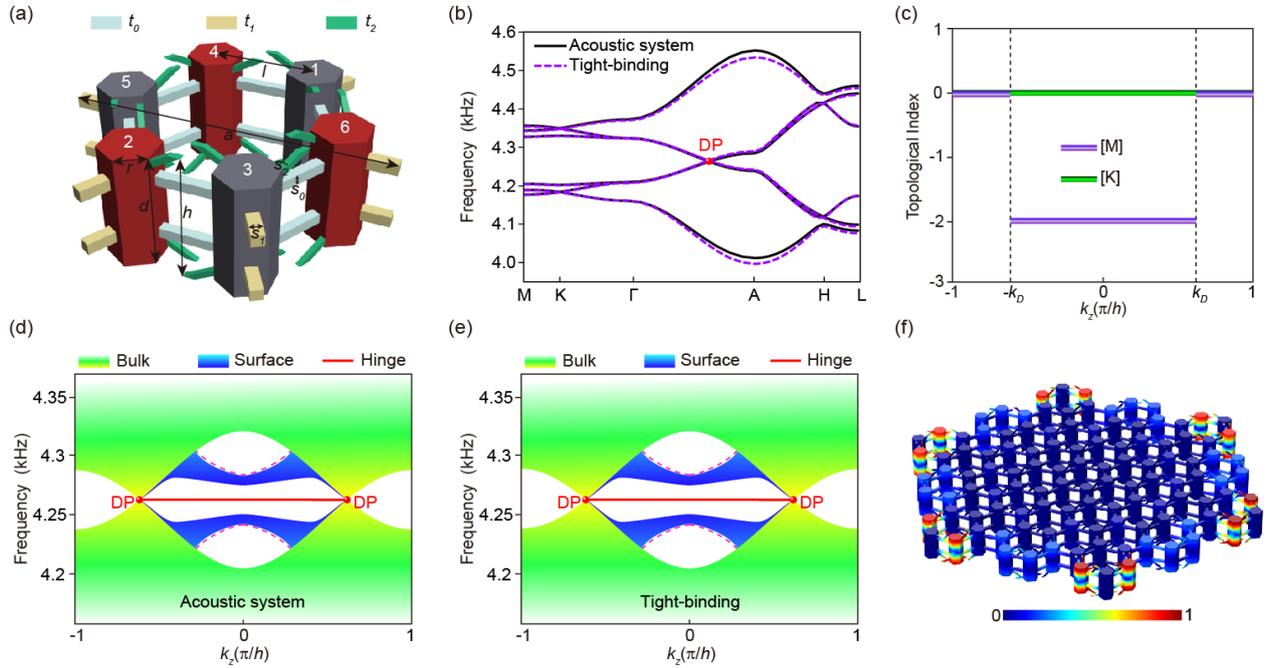

FIG. S3. Direct acoustic emulation of the 3D HO Dirac semimetal. (a) Unit cell structure. Physically, the air-filled cavity resonators emulate atomic orbitals, and the narrow tubes introduce couplings between them. (b) Bulk band structure of the acoustic cavity-tube system, compared with the one fitted by the 3D tight-binding model. The red sphere highlights the Dirac point. (c) Topological index plotted as a function of $k_z$, indicating nontrival $C_2$-invariant ($[M] = -2$) within the momentum range $|k_z| < |k_D|$. (d) Hinge spectrum simulated for the real acoustic structure, compared with the tight-binding model result in (e). In addition to the HO hinge states (red solid lines) pinned at zero energy, the system also exhibits topologically trivial hinge states

(pink dashed lines). (f) Simulated eigenfield profile of the hinge modes at $k_z = 0$ for the acoustic system, demonstrated with a small sample for clarity.

Here we use an acoustic cavity-tube strcuture to emulate the 3D tight-binding model directly. As shown in Fig. S3(a), the six (regular hexagonal prism) cavities mimic the orbitals, and the narrow (square) tubes emulate the couplings among them. Specifically, the in-plane and out-of-plane lattice constants are $a = 111$ mm and $h = 45$ mm, respectively. The cavity parameters $r = 10$ mm and $d = 40$ mm are selected to ensure the frequency of the $P_z$ dipole mode ($f_0 = 4264.5$ Hz) to be far away from other cavity modes, serving as a good single-mode approximation. This gives the constant onsite energy. The distance between the centers of two neighboring cavity resonators is $l = 37$ mm. Both the intra-cell coupling $t_0$ and the in-plane inter-cell coupling $t_1$ are emulated by one pair of horizontal square tubes (located at a distance of $d/2$), with the side-lengths $s_0 = 3.5$ mm and $s_1 = 4.0$ mm. The interlayer coupling $t_2$ is introduced by the inclined tube of side-length $s_2 = 2.1$ mm. As shown in the Fig. S3(b), the simulated band structure exhibits clearly fourfold degenerate Dirac points in the ΓA direction. The result is well fitted by the tight-binding model with couplings $t_0 = -56.5$ Hz, $t_1 = -73.5$ Hz, $t_2 = 20.5$ Hz, and $f_0 = 4264.5$ Hz. Note that the signs of the couplings are determined by the connectivity of the tubes between the cavities, negative for straight links and positive for inclined links (see Ref. 3 for details). With these couplings, we can derive a nontrivial $C_2$-invariant ($[M] = -2$) within the momentum range $|k_z| < |k_D|$, as shown in Fig. S3(c). This scenario is complementary to the 3D HO Dirac semimetal considered in Fig. 1 (see the main text), where the nontrivial $C_2$-invariant occur within the momentum range $|k_z| > |k_D|$. To identify the HO band topology, we simulate the hinge-projected spectrum for an infinitely long hexagonal prism. As shown in Fig. S3(d), the HO hinge states, which are flat and pinned at the resonant frequency of the cavity, emerge and connect the pair of projected Dirac points within $|k_z| < |k_D|$, consistent with the $k_z$-dependent $C_2$-invariant. In addition to the HO hinge states, the hinge spectrum also exhibits topologically trivial (gapped) surface modes and hinge modes. All these features precisely reproduce the hinge spectrum calculated for the tight-binding model, as shown in Fig. S3(e). (Note that in contrast to the hinge modes emerging within $|k_z| > |k_D|$ in the main text, here a new set of parameters are used, and the hinge modes appear within $|k_z| < |k_D|$.) The Fig. S3(f) demonstrates the eigenfield profile of the hinge modes simulated for the acoustic system, which is highly-confined to the corners of the sample.

In principle, the system can be fabricated with the 3D printing technique. However, there are two difficulties that hinder the experimental characterization of the sample.

(i) It is very hard to measure the bulk states since one cannot insert the microphone into the inner cavities of the sample.

(ii) It is not easy to resolve the hinge states in frequency, given the narrow frequency window of the hinge states with respect to the frequency of interest (~2%, see Fig. S2d).

### 4. Numerical and experimental methods

*Numerical simulations.* All full-wave simulations were performed by using a commercial solver package (COMSOL Multiphysics). The photosensitive resin material used for fabricating samples was modeled as acoustically rigid in the airborne sound environment, given the extremely mismatched acoustic impedance between resin and air. The air density 1.29 kg/m³ and the sound speed 343 m/s were used to solve the eigen-problems in Figs. 2(b) and 2(c). Specifically, the bulk band structure in Fig. 2(b) was obtained by imposing Bloch boundary condition along all directions. The hinge spectrum in Fig. 2(c) was simulated for an infinitely long hexagonal prism, i.e., with rigid boundary condition for its side surfaces and Bloch boundary condition along the $z$ direction. Similar treatments are made for the bulk and hinge spectra in Fig. S3.

*Experimental measurements.* Our experimental sample [Fig. 3(a)], which has the geometry of a regular hexagonal prism, was fabricated via the 3D printing technique with a nominal fabrication error ~0.2 mm. To detect the bulk information, a broadband point-like sound source, launched from a subwavelength-sized tube, was directly inserted into the middle of the sample from its top facet, and a 1/4 inch microphone (B&K Type 4187) was used to scan the pressure information through a vertical hole adjacent to the middle one [see the left panel in Fig. 3(b)], with a spatial step of $h = 9.6$ mm. Both the amplitude and phase information of the input and output signals was recorded and frequency-resolved with a multi-analyzer system (B&K Type 3560B). The bulk spectrum in Fig. 3(c) was obtained by performing 1D Fourier transform of the measured data. Special treatment was implemented to detect the hinge modes that are strongly localized at the hinges [see the inset of Fig. 2(c)]. Specifically, the cover plate was perforated with 40 equidistant side holes (of radius 0.4 mm) in the vicinity of a sample hinge. The point source was fixed in the middle side hole, and the microphone was inserted into the side holes and moved to scan the pressure information along the hinge [see the right panel in Fig. 3(b)]. The side holes not in use were sealed during measurements. The hinge spectrum in Fig. 3(d) was obtained by performing 1D Fourier transform of the measured data. To demonstrate the field distributions in Fig. 3(e), additional data measured from other vertical holes were supplemented.